# A NOVEL (K,N) SECRET SHARING SCHEME FROM QUADRATIC RESIDUES FOR GRAYSCALE IMAGES

El-Tigani B. Abdelsatir, Sahar Salahaldeen, Hyam Omar and Afra Hashim
Department of Computer Science, Faculty of Mathematical Sciences
University of Khartoum, Sudan

## ABSTRACT

*A new grayscale image encryption algorithm based on $(k, n)$ threshold secret sharing is proposed. The scheme allows a secret image to be transformed into n shares, where any $k \leq n$ shares can be used to reconstruct the secret image, while the knowledge of $k - 1$ or fewer shares leaves no sufficient information about the secret image and it becomes hard to decrypt the transmitted image. In the proposed scheme, the pixels of the secret image are first permuted and then encrypted by using quadratic residues. In the final stage, the encrypted image is shared into n shadow images using polynomials of Shamir scheme. The proposed scheme is provably secure and the experimental results shows that the scheme performs well while maintaining high levels of quality in the reconstructed image.*

## 1. INTRODUCTION

The use of digital images to communicate has increasingly become a new way of communication nowadays especially in military, medical and financial applications. Therefore, protecting digital images containing sensitive or confidential information has become an important issue. Encryption is a widely used technique to achieve information security goals by transforming data into unreadable format. However, in comparison with textual information, image data is different in properties, such as the high correlation of image pixels and high redundancy. Therefore, the application of conventional methods of encryption in digital images is not recommended. Moreover, secret image sharing eliminates weaknesses in pure image encryption. For example, if the secret key in encryption is lost the entire secret image becomes invalid. Also, in situations where media failure is more common, there is a risk if the secret image is held by only one person, the secret data may be lost or modified. This weakness can be avoided in $(k, n)$ threshold secret sharing schemes in which the image is encoded into n shadow images and only $k \leq n$ of them can reconstruct the original image.

The concept of secret sharing scheme was originally proposed by Shamir [1] and Blakely [3] independently in 1979 for textual information. The research of secret sharing in digital images first appeared in 1994 by the work of Naor and Shamir [2] in which the human visual system is used directly to recover the secret image and the scheme was computationally efficient.

In recent years, along with the prevalent advancements in image processing, secret image sharing also has been in active research. A wide variety of methods have been developed for processing grayscale images [9] [13] [10], and also for colour images [6], [7],[8]. While many proposals follow the classical idea of Shamir secret sharing scheme with variations, in some schemes, the size of the shared images is much bigger than the original one. Moreover, a common disadvantage of existing schemes is that there is a great quality loss between the secret image and the recovered one [4] [12].

In this paper, we extend (2,2) Chen-Chang method [5] to $(k, n)$ secret sharing threshold scheme and as a result we propose a new scheme for secret image sharing from quadratic residues and incorporated with Shamir secret sharing method [1]. The proposed scheme is provably secure and prevents the secret image from being revealed by making use of quadratic residues to encrypt the secret image and then the encrypted pixels are shared into $n$ image shares by Shamir secret sharing method.  The scheme provides an advantage to Chen-Chang method [5] in terms of flexibility in the number of secret shares which is limited to only two shares. Moreover, the experimental results show nearly identical quality of the reconstructed image compared with the original one. Also, the size of the secret image and the image shares is the same as the size of the original secret image. Additionally important, the new scheme provides a better security measures to protect the secret image in comparison with [5] and [4].

The rest of this paper is organized as follows. In Section 2, we review the schemes of Shamir, Naor-Shamir, Thien-Lin and Chen-Chang. In Section 3, we propose efficient $(k, n)$ threshold method for secret image sharing. In Section 4, we give experimental results and finally in Section 5 we draw our conclusions.

## 2. RELATED WORK

### 2.1. Shamir Secret Sharing

We begin by introducing the original Shamir's scheme [1]. Suppose that we want to encode the secret S into n shares $(S1, S2, ... , Sn)$ and we wish that the secret data S cannot be revealed without k or more shares. In Shamir secret sharing scheme the partition of the secret is done by the following polynomial:

$$F(x_i) = y + m_1 x_i + m_2 x_i^2 + \cdots + m_{(k-1)} x_i^{(k-1)} \bmod(p), \qquad i = 1,2, \dots, n \qquad (1)$$

where y is the share, $S_1$ , p is a prime number and the cofficients of the $k - 1$ degree polynomial $m_i$ are chosen randomly and then the shares are evaluated as $S1 = F(1), S2 = F(2), ... , Sn = F(n)$

Given any k pairs of the share pairs $\{(i, Si)\}, i = 1,2.., n.$ we can obtain the coefficients $m_i$ of F(x) by largrange interpolation as follows:

$$S = (-1)^{(k-1)} \left[ F(x_1) \frac{(x_2)(x_3)\dots(x_k)}{(x_1-x_2)(x_1-x_3)\dots(x_1-x_k)} + F(x_2) \frac{(x_1)(x_3)\dots(x_k)}{(x_2-x_1)(x_2-x_3)\dots(x_2-x_k)} + \cdots + F(x_k) \frac{(x_1)(x_2)\dots(x_{k-1})}{(x_k-x_1)(x_k-x_2)\dots(x_k-x_{k-1})} \right] \qquad (2)$$

### 2.2. Naor-Shamir Scheme

In 1994 Naor and Shamir [2] proposed the first secret image sharing scheme in which a secret monochrome image is encrypted into two shares. Monochrome pixels have only two values either black or white. The pixels of the two shares are determined with probability 50%. If the pixel is white, one of the above two rows is chosen to generate two shares. Similarly If pixel is black, one of the below two rows of is chosen to generate the two shares. Figure 1. shows the possible values of pixels in each of the two generated shares.

| Pixel | Probability | Share1 | Share2 | Share1 Share2 |
|---|---|---|---|---|
|  | 50% |  |  |  |
|  | 50% |  |  |  |
|  | 50% |  |  |  |
|  | 50% |  |  |  |

Figure 1 : Naor-Shamir Scheme

The reconstruction of the secret image is simply done by stacking the pixels again as shown in Figure 1. Although this scheme eliminates the overhead of complex computation present in Shamir's scheme, however, the scheme is valid for monochrome images only and the secret image can be detected and reconstructed by the human visual system with no much effort.

### 2.3. Thien-Lin $(k, n)$ Scheme

Thien and Lin [4] proposed a lossy (k, n) threshold image secret sharing scheme derived from Shamir secret sharing method. The scheme generates small n shadows of size rxr from polynomials of degree $k - 1$. In this scheme, the pixels of greater than 250 are truncated and then permuted with a secret key.

Each time k pixels are proccesed from the permuted set of pixels and a polynomial of degree $k - 1$ is created.

$$F(x_i)j = p_0 + p_1 x_i + p_2 x_i^2 + \cdots + p_{(k-1)} x_i^{(k-1)} \mod 251, \qquad i = 1,2, \ldots, n \qquad (3)$$

Where $m_i$ are values from the set of permuted pixels and $j$ is the current iertation $1 \leq j, \leq r$ .The $n$ shares are calculated as $F_j(1), F_j(2), F(3), \ldots, F_j(n)$, and sequentially assigned to the n shadow images $S_1, S_2, S_3, \ldots, S_n$. The algorithm is repeated until all permuted pixels are processed. The revealing process is implemented by Lagrange interpolation using any k shadow images. For each j value, $f_j(1), f_j(2), \ldots, f_j(k)$ are computed to obtain the coefficients $p_0, p_1, p_2, \ldots, p_{k-1}$ by interpolation then the inverse of the permutation operation is applied to recover the original secret image.

### 2.4. Quadratic Residues and Chen-Chang $(2, 2)$ scheme

Let $p \in Z$ and $p \in N$ . A number $x$ is called a quadratic residue (QR) modulo $p$ if $gcd(x, p) = 1$ where gcd is the common greatest divisor for $x$ and $p$, and the congruence

$$y^2 \equiv x \bmod p \qquad (4)$$

has a solution and x is called a quadratic nonresidue (QNR) $modulo\ p$ if (4) has no solution.

Let p and q be two prime numbers satisfy that $(p + 1 \bmod 4 = 0)$, $(q + 1 \bmod 4 = 0)$ and $n = p.q$. The sets of QR and NQR with respect to $p$ and $q$, Respectively, can be computed as follows:

$$S_{QR-P} = \{1^2, 2^2, \ldots, \left(\frac{p-1}{2}\right)^2 \bmod p\} \tag{5}$$
$$S_{NQR-p} = \{x: 0 < x < p, for\ any\ integer\ x\} - S_{QR-P}$$
$$S_{QR-q} = \{1^2, 2^2, \ldots, \left(\frac{q-1}{2}\right)^2 \bmod q\}$$
$$S_{NQR-q} = \{x: 0 < x < q, for\ any\ integer\ x\} - S_{QR-q}$$

Now, for the $x$ there are four cases:

Case 1: $x$ is QR to both $p$ and $q$.
Case 2: $x$ is QR to $p$, but $x$ is QNR to $q$.
Case 3: $x$ is QNR to $p$, but $x$ is QR to $q$.
Case 4: $x$ is QNR to both $p$ and $q$.

$x$ must be a QR to both $p$ and $q$, that is done by multiplying corresponding parameter $r \in \{r_1, r_2, r_3, r_4\}$ where $r$ is the result from intersection of the sets of QR and QNR with respect to $p$ and $q$ which is defined as follows:

Case 1: $r_1 \in R_1 = S_{QR-P} \cap S_{QR-q}$.
Case 2: $r_2 \in R_2 = S_{QR-P} \cap S_{NQR-q}$.
Case 3: $r_3 \in R_3 = S_{QNR-P} \cap S_{QR-q}$.
Case 4: $r_4 \in R_4 = S_{QNR-P} \cap S_{QNR-q}$.

Once $x$ is QR to both $p$ and $q$ then it has four square roots namely $M_1, M_2, \ldots, M_4$ and they can be computed by the following equations:

$$\begin{aligned} M_1 &= (a.m_1 + b.m_3) \bmod n \\ M_2 &= (a.m_1 + b.m_4) \bmod n \\ M_3 &= (a.m_2 + b.m_3) \bmod n \\ M_4 &= (a.m_2 + b.m_4) \bmod n \end{aligned} \tag{6}$$

Where:
$$\begin{aligned} m_1 &= x^{(p+1)/4} \bmod p \\ m_2 &= (p - m_1) \bmod p \\ m_3 &= x^{(q+1)/4} \bmod q \\ m_4 &= (q - m_3) \bmod q \\ a &= q.(q^{-1} \bmod p) \\ b &= p.(p^{-1} \bmod q) \end{aligned} \tag{7}$$

The sharing and reconstruction phases of Chen-Chang scheme are described as follows:

**The sharing phase:**

Let $p = 239$ and $q = 251$, which are the greatest two prime numbers not larger than 255. Decrease the value of each pixels in the image higher than 238 to 238, Compute the modified pixel $\bar{p}_i = p_i.r \bmod n$ for each pixel such that $\bar{p}_i$ is a QR to both $p$ and $q$. Now $\bar{p}_i$ has four square roots modulo $n$ namely $M_1, M_2, M_3$ and $M_4$ then choose one of them $M = M_t$ according to $r = r_t$ $t \in \{1,2,3,4\}$. The last phase is to split $M$ into two bytes $a_i$ and $b_i$, put each of it in a shadow images $I_a$ and $I_b$ respectively.

**The reconstruction phase:**

Once the shadow images $I_a$ and $I_b$ are received, take each pixel from both images and concatenate them to get the square root $M'$ ($M' = a_i' \| b_i'$) . Now compute $p_i' = M'^2 \bmod n$ then find the four square roots modulo $n$ for $p_i'$ namely $M_1', M_2', M_3'$ and $M_4'$ . Finally, choose the correct parameter $r_t$ based on the corresponding $M_t'$ where $t \in \{1,2,3,4\}$ and get the original image pixel by $p_i = \frac{p_i'}{r} \bmod n$.

## 3. THE PROPOSED SCHEME

In this section we propose a new $(k,n)$ secret image sharing scheme based on Chen-Chang scheme and Shamir's method to encode an image $S$ to $n$ shadow images, $S1, S2, S3, \ldots, Sn$. In a way similar to [5], we use quadratic residues to encrypt image pixels, however, Chen-Chang's scheme is a (2, 2) secret sharing scheme and the two shares are obtained and recovered by a simple split and concatenation operation . If the two participants in the algorithm combine their shares by the reveal algorithm, the secret is discovered . As a result, the sharing procedure used in [5] is not suitable for the general $(k,n)$ threshold scheme. In this work, we extend the scheme by Chen-Chang and replace the split and concatenation operation by using Shamir's method of secret sharing. The sharing and reveal steps of the proposed scheme are described below.

**The Sharing Algorithm:**

Input: the secret image S
Output: n shadow images

1- Decrease pixel value $p_i$ if it is greater than 238 to 238.
2- Shuffle the pixels of the original image S to obtain a permuted set of pixels O.
3- Choose the proper case for the pixel $p_i$ of O.
4- Choose the first element $r$ in a proper $R_i$ where i = 1,2,3,4.
5- Compute the product $\bar{p}_i = p_i.r$ .
6- Find the four square roots for $\bar{p}_i$ .
7- Choose $M = M_i$ according to $r_i$ .
8- Use Shamir secret sharing encoding to transform $M$ into $n$ shares.
9- Construct $n$ shadow images from the $n$ shares.

**The Reveal Algorithm:**

Input: $k <= n$ shadow images.
Output: the secret image S.

1. Perform the polynomial interpolation of Shamir decoding algorithm to get $M'$
2. Compute $\bar{p}_i = M'^2 \bmod n$ .
3. Find the four square root of $\bar{p}_i$ .
4. Find appropriate parameter $r'$ according to the order of $M'$.
5. Compute $p_i = \bar{p}_i/r'$ .
6. Put $p_i$ in the reconstructed image.
7. After all pixels $p_i$ are processed, Depermute the reconstructed O to get S.

The proposed scheme is provably secure since it follows Shamir's scheme and quadratic residues encryption which relies on the hardness of the composite quadratic residuosity problem defined as: Let $= pq$, $m \epsilon \mathbf{Z}_n$ and the $gcd(m,n) = 1$ where p and q are unknown primes. It is computationally hard to determine whether $m \epsilon QR(n)$ or $m \epsilon QNR(n)$.

**Example:**

Let $p_i = 150$, n=5, and k=3, we find that $p_i$ is of case 2 and $r = r_2 = 2$, so $\bar{p}_i = 150.2 = 30$. The four square roots modulo 59989 are $M_1 = 42677, M_2 = 52954, M_3 = 7035, and\ M_4 = 17312$, and we choose $M = M_2 = 52954$ for that $p_i$ is of case 2.

When applying Shamir's sharing algorithm on M, The resulting polynomial is $(x) = 52954 + 173x + 167x^2$. In this example four shares in the form (x, f(x)) are calculated where $x \in \{1,2,3,4,5\}$.

M is transformed into four pixels: $a_1 = 236, a_2 = 193, a_3 = 6, a_4 = 153$ and shared into 4 shadow images $I'_1, I'_2, I'_3, I'_4$ respectively.

When the receiver gets $a_1, a_2, a_3, a_4$ from shadow images $I'_1, I'_2, I'_3, I'_4$ respectively and after Lagrange interpolation is calculated we get $M' = 52954$. Then $\overline{p'_i} = {M'}^2 mod\ n = 300$ and four square roots modulo n of $\overline{p'_i}$ are $42677, 52954, 7035, and\ 17312$ since $M'$ is in the 2$^{nd}$ order of four square roots, $r = r2 = 2$ and $p'_i = \frac{\overline{p'_i}}{r'} mod\ n = 150$.

## 4. EXPERIMENTAL RESULTS

The proposed scheme shows great performance in terms of low distortion in the reconstructed image. In Figure 2, five well-known grayscale images of the size 512x512 are used for the experiment. We use Peak Signal to Noise Ratio (PSNR) an image quality metric to measure the difference between the secret images and the recovered ones. PSNR is defined as follows:

$$MSE = \frac{1}{w*h} \sum_{i=1}^{w*h}(\widetilde{y_i} - y_i)^2 \quad PSNR = \log_2 {255^2}/{MSE} \qquad (8)$$

Where $\widetilde{y_i}$ is the pixel i in the reconstructed image, $y_i$ is the pixel of the original image and MSE (Mean Square Error) is the difference between the two images. High value of PSNR indicates higher quality of image.

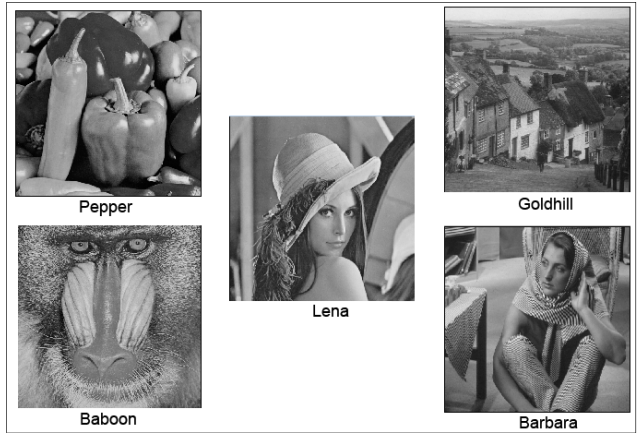

Figure 2: Test images

In our experiment a (2,4) secret sharing scheme was chosen. i.e. Here n=4, k=2. Figure 3 shows the secret image (Lena), the shares and the reconstructed image. Table 1 lists the experimental result. In four cases, ∞ means that the proposed scheme give identical reconstructed image as the original one and it is impossible to distinguish the secret image from the reconstructed one. According to [11], the acceptable range of a lossy image is between 30 and 50 dB and only in one case, the obtained value 0.93972 dB is much higher than the average.

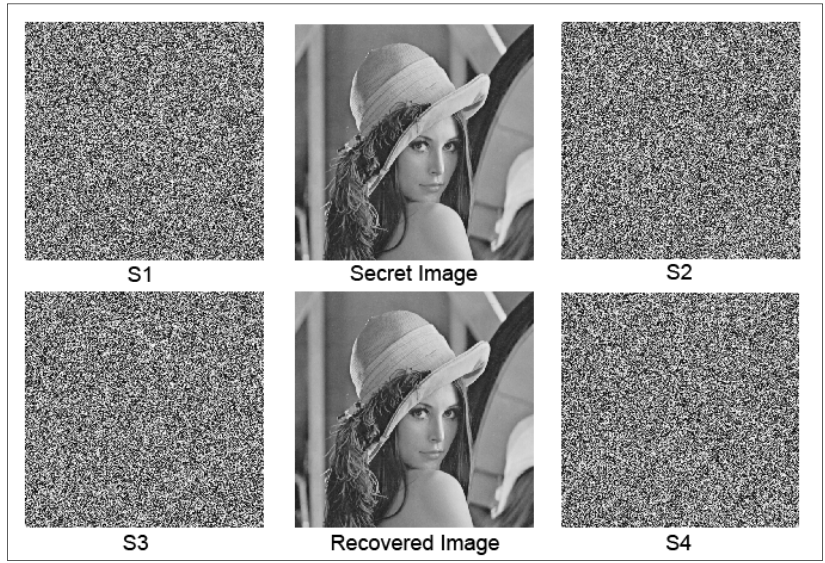

Figure 3: The secret image, recovered image and the four noise-like image shares $S1, ..., S4$

Table 1. PSNR Values.

| Secret Image | Proposed Method N=4, k=2 |
|---|---|
| Peppers | ∞ |
| Lena | 93.97 dB |
| Baboon | ∞ |
| Barbara | ∞ |
| Goldhill | ∞ |

## 5. CONCLUSION

In this paper a new $(k, n)$ threshold secret sharing method for grayscale images is introduced. The concept of quadratic residues is used to encrypt the image and then Shamir's approach is employed to produce the shadow images. The main drive of this work is to allow any number of participants to share a secret image in a secure manner. The proposed scheme provides greater flexibility and security in comparison with Chen-Chang's method. The experimental results shows that advantages of our scheme are: (i) Identical or nearly similar reconstructed image from any *k* shadow images, (ii) k shadow images with the same size as the original image. Our future work will investigate the implementation of the proposed scheme effectively in embedded and mobile devices.

## REFERENCES

[1]     Adi Shamir (1979) How to share a secret, Commun. ACM, vol 22, No. 11, pp612-613.
[2]     Moni Naor & Adi Shamir (1995) Visual cryptography, Advances in Cryptology — EUROCRYPT'94 Lecture Notes in Computer Science, Vol 950, pp1-12.
[3]     George Blakley (1979) Safeguarding Cryptographic Keys, Proceedings of the National Computer Conference, pp313-317.
[4]     Chih-Ching Thien & Ja-Chen Lin (2002) Secret Image Sharing, Computers and Graphics, Vol 26, pp765-770.
[5]     Chang-Chu Chen & Chin-Chen Chang (2007) Secret image sharing using quadratic residues, Intelligent Information Hiding and Multimedia Signal Processing (IIHMSP), Vol 1, pp515-518.
[6]     Rastislav Lukac & Konstantinos Plataniotis (2004) A color image secret sharing scheme satisfying the perfect reconstruction property, 6th Workshop on Multimedia Signal Processing, pp351-354.
[7]     Shyamalendu Kandar & Bibhas C Dhara (2012) Random sequence based secret sharing of an encrypted color image, 1st International Conference on Recent Advances in Information Technology (RAIT), pp.33-37.
[8]     Chin-Chen Chang & Iuon-Chang Lin (2003) A new (t, n) threshold image hiding scheme for sharing a secret color image, International Conference on Communication Technology Proceedings, Vol 1 ,pp.196-202.
[9]     Prabir Naskar , Ayan Chaudhuri, Debarati Basu & Atal Chaudhuri (2011) A Novel Image Secret Sharing Scheme, Second International Conference on Emerging Applications of Information Technology (EAIT), pp177-180.
[10]    Runhua Shi , Zhong Hong, Liusheng Huang & Yonglong Luo (2008) A (t, n) Secret Sharing Scheme for Image Encryption, Congress on Image and Signal Processing, Vol 3, pp3-6.
[11]    Mauro Barni (2006) Document and Image Compression, CRC Press.
[12]    Chien Ming-Chun & J.G. Hwang (2012)  Secret image sharing using (t,n) threshold scheme with lossless recovery,5th International Congress on Image and Signal Processing (CISP), pp1325-1329.
[13]    Chang-Chou Lin & Wen-Hsiang Tsai (2003) Visual cryptography for gray-level images   by dithering techniques, Pattern Recognition Letters, Vol 24, No. 1–3, pp349-358.